\pdfoutput=1
\documentclass[aps,prb,twocolumn,superscriptaddress,groupedaddress,floatfix,longbibliography]{revtex4-1}  

\usepackage{graphicx}  
\usepackage{dcolumn}   
\usepackage{bm}        
\usepackage{amssymb}   
\usepackage{amsmath}   
\usepackage{amsthm}    
\usepackage{xcolor}    
\usepackage{subfigure}
\usepackage{soul}

\definecolor{cbblue}{RGB}{109,182,255}
\definecolor{cbyellow}{RGB}{255,255,109}
\definecolor{cbpurple}{RGB}{73,0,146}

\begin{document}


\title{
Multi-faceted machine learning of 
competing orders in disordered interacting systems
}
       
\author{Michael Matty$^1$}
\email{mfm94@cornell.edu}
\author{Yi Zhang$^1$}
\author{Zlatko Papi\'c$^2$}
\author{Eun-Ah Kim$^1$}
\email{eun-ah.kim@cornell.edu}

\affiliation{$^1$Department of Physics, Cornell University, Ithaca, New York 14853, USA}
\affiliation{$^2$School of Physics and Astronomy, University of Leeds, Leeds LS2 9JT, UK}

\date{\today}

\begin{abstract}
While the non-perturbative interaction effects in the fractional quantum Hall regime can be readily simulated through exact diagonalization, it has been challenging to establish a suitable diagnostic that can label different phases in the presence of competing interactions and disorder.
Here we introduce a multi-faceted framework using a simple artificial neural network (ANN) to detect defining features of a fractional quantum Hall state, a charge density wave state and a localized state using the entanglement spectra and charge density as independent input.
We consider the competing effects of a perturbing interaction ($l=1$ pseudopotential $\Delta V_1$), a disorder potential $W$, and the Coulomb interaction to the system at filling fraction $\nu=1/3$. Our phase diagram benchmarks well against previous estimates of the phase boundary using conventional measures along the $\Delta V_1=0$ and $W=0$ axes, the only regions where conventional approaches have been explored. Moreover, exploring the entire two-dimensional phase diagram for the first time, we establish the robustness of the fractional quantum Hall state and map out the charge density wave micro-emulsion phase wherein droplets of charge density wave region appear before the charge density wave is completely disordered.  
Hence we establish that the ANN can access and learn the defining traits of topological as well as broken symmetry phases using multi-faceted inputs of entanglement spectra and charge density.

\end{abstract}

\maketitle

\newpage

\section{Introduction}

Using the fact that neural network based
machine learning can effectively distill relevant information and compactly represent a
complex function, there have been recent efforts to efficiently (i) obtain phase
diagrams
~\cite{broecker:sr2017a,broecker:ap2017a,zhang:prl2017a,zhang:pr2017a,wang:pr2016a,carleo:s2017a,carrasquilla:np2017a,nieuwenburg:np2017a,beach:pr2018a,chng:pr2017a,chng:pr2018a,deng:pr2017b,liu:prl2018a,nieuwenburg:pr2018a,ohtsuki:jpsj2016a,schindler:pr2017a,wetzel:pr2017b,wetzel:pr2017a,yoshioka:pr2018a,venderley:prl2018a}
and (ii) represent wave functions
~\cite{cai:pr2018a,carleo:s2017a,chen:pr2018a,deng:pr2017a,deng:pr2017b,gao:nc2017a,huang:ap2017a,liu:pr2017b,nomura:pr2017a,schmitt:sp2018a,torlai:np2018a} of many-body quantum systems. Machine learning based phase detection is a particularly promising
direction for phases outside the traditional ''knowledge compression scheme'' -- the local
order parameter. Topological phases form a central class of such phases.
Though there has been recent progress in using machine learning for topological phases
~\cite{beach:pr2018a,deng:pr2017b,sun:pr2018a,yoshioka:pr2018a,ohtsuki:jpsj2016a,zhang:pr2017a,zhang:prl2017a,zhang:prl2018a},
these early efforts naturally centered around benchmarking the neural
network based approaches to the conventional approaches on established
problems by suppressing either disorder or interaction.

Here we turn to a strongly interacting two-dimensional electron gas (2DEG) in  the fractional quantum Hall regime. 
2DEG hosts a rich phase diagram in which topological order in various quantum Hall states competes against various forms of spontaneously broken symmetries. All this can only be observed in clean samples, which abundantly speaks to the key role of disorder. 
Such competition among interaction driven correlated states and disorder driven localized states  is a common theme of all strongly correlated systems. Yet,  2DEG forms the simplest system in which such competition can be studied experimentally, with the magnetic field quenching the kinetic energy.
Nevertheless theoretical study is challenging since 
the traditional diagnostic of the fractional quantum Hall state requires translational symmetry.

Previous works~\cite{sheng:prl2003a,wan:pr2005a,liu:pr2017a,liu:prl2016a}
on the effect of disorder on the FQH effect focused on establishing a measure that can assess the robustness of the topological order. This was already a challenging task because most measures of topological order are naturally suited to uniform systems with translation invariance. The resulting measures of the total Chern number\cite{sheng:prl2003a,wan:pr2005a} and the slope of the entanglement entropy as a function of average disorder strengths
\cite{liu:pr2017a,liu:prl2016a} are compelling. Nevertheless, they are computationally costly. More importantly, these measures are mostly geared towards a bi-partite phase diagram consisting of a FQH state and a localized, insulating state (for example, the quantization of the Chern number would not distinguish between different incompressible phases realized at the same filling factor). 
Here we study the problem including \emph{both} disorder and a competing interaction from two complementary perspectives: the entanglement spectrum (ES) and the real space charge density (CD).

Motivated by the fruitful use of entanglement spectra in clean systems with topological order \cite{li:prl2008a,thomale:prl2010a,qi:prl2012a}, efforts sought indication of the robustness of the topological state within the entanglement spectrum in the presence of disorder
~\cite{prodan:prl2010a,liu:pr2017a,liu:prl2016a}.
The ES $\{ \xi_i \}$ is a set of eigenvalues of the reduced density matrix, characterizing a subsystem of a quantum system~\cite{li:prl2008a}.
Traditionally, the levels in the ES are organized according to a symmetry quantum number (typically, linear or angular momentum), which reveals characteristic structures in translationally invariant systems. However, such an organizing principle is lost with the introduction of disorder, and one must resort to studying the \emph{distribution} of the entanglement spectrum levels as previously used, e.g., in the studies of many-body localization~\cite{geraedts:ap2016a, Serbyn2016}.
Alternatively, the real space charge density provides a two-dimensional image of the wave function that can characterize phases through its spatial profile (for example, the charge
density is uniform in the FQH liquid state).  
However, in the presence of interactions it becomes non-trivial to  assign phase boundaries simply from 
images of the charge density. 
Thus, competing interactions and disorder necessitate
an approach that can discover distinguishable structure in multiple facets: the ES and the CD.

In this paper we use supervised machine learning on  ES and CD to
obtain the phase diagram of three competing phases tuned by interaction and disorder strength: FQH, charge density wave (CDW), and a localized state.   This approach is versatile and numerically efficient. It can be generalized to incorporate other phases, system geometries, and disorder models. We start by 
briefly reviewing the standard theoretical model for the FQH system in the presence of Gaussian white noise disorder in Sec.~\ref{sec:model}. In Sec.~\ref{sec:method} we introduce our method based on ANN. Our results are presented in Sec.~\ref{sec:results}. We conclude with a summary of our results and open questions in Sec.~\ref{sec:conc}.  

\section{Model}\label{sec:model}

\begin{figure}[htb]
    \centering
    \includegraphics[width=0.7\linewidth]{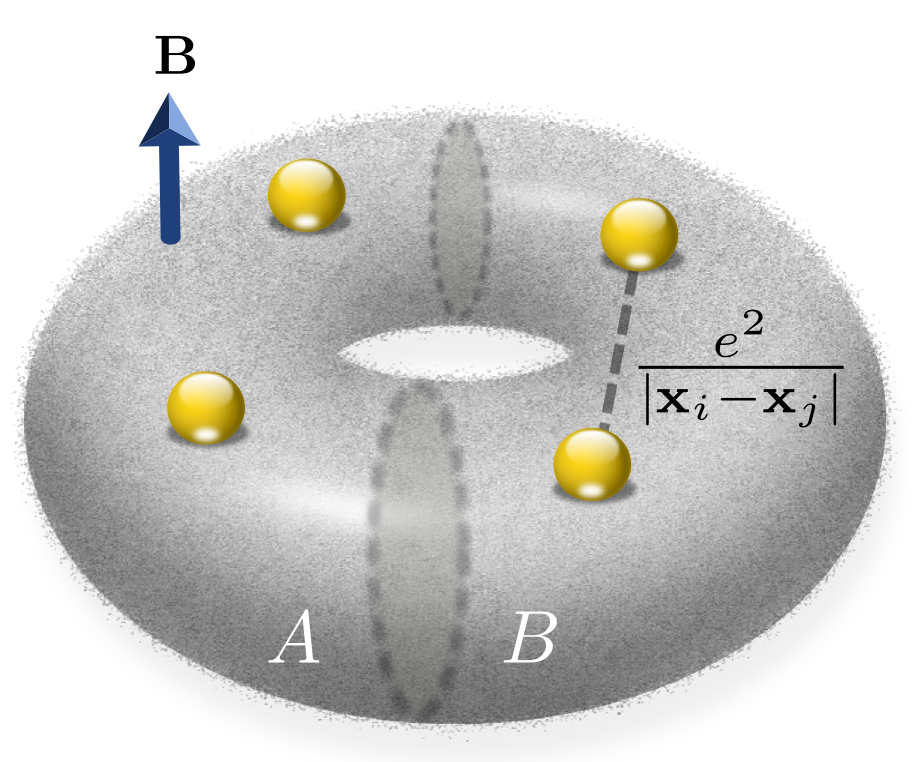}
    \caption{Two-dimensional electron gas on a torus with perpendicular magnetic field $B$. Electrons interact via the Coulomb potential in Eq.~(\ref{eq:ham0}), while the grainy surface of the torus depicts the presence of Gaussian white noise disorder, Eq.~(\ref{eq:dis}). The partitioning of the system into parts $A$ and $B$, used to define the entanglement spectrum, is also indicated.
    }
    \label{fig:torus}
\end{figure}

We consider a system of electrons on a torus in the presence of a magnetic field, see Fig.~\ref{fig:torus}. The area of the torus $\mathcal{A}$ is set by the magnetic
flux, 
$\mathcal{A}=2\pi\ell_B^2 N_\phi$
in units of the magnetic area $2\pi\ell_B^2$, where $\ell_B=\sqrt{\hbar/eB}$ is the magnetic length. We set the aspect ratio of the torus to be unity, corresponding to a square unit cell with periodic boundary conditions.
The electron density is held fixed at one electron per three magnetic fluxes, i.e., the filling fraction is $\nu=1/3$. 

The system is described by the Hamiltonian
\begin{eqnarray}\label{eq:ham}
H=H_{0}+ H_{\rm pert},
\end{eqnarray}
where $H_0$ represents the Coulomb interaction between the electrons,
\begin{eqnarray}\label{eq:ham0}
H_0 = \sum_{i<j}\frac{e^2}{|\mathbf{x}_i-\mathbf{x}_j|},
\end{eqnarray}
and the $\mathbf{x}_i$'s denote the positions of the electrons in the 2D plane. 
We emphasize that all the terms in the Hamiltonian are explicitly projected to the lowest Landau level using standard techniques~\cite{fano:pr1986a,prange:s1990a}. Physically, this corresponds to taking the limit of an infinite magnetic field, which is an excellent approximation for most purposes~\cite{prange:s1990a}.
In addition to the Coulomb potential, we consider two physical perturbations: 
\begin{eqnarray}
H_{\rm pert} = H'(\Delta V_1)+ H_{\rm imp}(W).
\end{eqnarray}
where $H'$ denotes the perturbation by a short-range (contact) interaction between the electrons, 
\begin{eqnarray}
H'(\Delta V_1) = \Delta V_1 \sum_{i<j}\nabla_i^2 \delta (\mathbf{x}_i - \mathbf{x}_j),
\end{eqnarray}
also known as the Haldane $V_1$ pseudopotential~\cite{haldane:prl1983a}. The strength of this perturbation is denoted by the overall prefactor $\Delta V_1$. 
Physically, this perturbation could arise due to effects of finite thickness of the 2D electron gas, excitations to other Landau levels, etc. 
On the other hand, the quenched disorder potential is denoted by a (one-body) term $H_{\rm imp}(W)$. 
We model disorder as Gaussian white noise~\cite{wan:pr2005a} randomly distributed with
mean value $0$ and width $W$ in real space, i.e.
\begin{eqnarray}\label{eq:dis}
     \langle H_{\rm imp}(W,\mathbf{x}) \rangle &=& 0,\\
     \langle H_{\rm imp}(W,\mathbf{x})H_{\rm imp}(W,\mathbf{x}') \rangle &=& 
    W^2 \delta(\mathbf{x}-\mathbf{x}').
\end{eqnarray}
The important feature for our purposes, which is shared with other common disorder models such as finite-range scatterers or correlated impurity
potentials, is that the disorder potential breaks magnetic translation invariance, thus we do not have a good quantum number to classify the many-body states.

We anticipate three phases in the parameter
space $(W,\Delta V_1)$ of the above model. First, we expect 
the Laughlin $\nu=1/3$ FQH state~\cite{laughlin:prl1983a} in the absence of $H'$ and $H_{\rm imp}$~\cite{haldane:prl1985a}. Second, 
decreasing the $V_1$ pseudopotential lowers the energy of the finite
wave vector magnetoroton excitation~\cite{GMP}, driving a transition to
a topologically trivial CDW state
at large enough $-|\Delta V_1|$~\cite{haldane:prl1985a}.
Finally, increasing the characteristic strength of the disorder $W$ eventually leads to a
localized state~\cite{wan:pr2005a}. 
Nevertheless, mapping out the phase diagram in the space of $(W,\Delta V_1)$  with these competing tendencies requires not only a numerical study but more importantly a new diagnostic.

\section{Method}\label{sec:method}

We use an artificial neural network (ANN) to diagnose structures in the ES and CD data characteristic of each phase.  In particular, we expect that the ANN will learn to distinguish
the universal part of the ES characterizing the topological FQH from the topologically 
trivial phases and the different entanglement entropy  of the
trivial phases distinguished by spontaneously versus explicitly broken continuous symmetry
\cite{metlitski:ap2015a}.  From the CD perspective, we expect the network to distinguish
zero (FQH), one (CDW), and two (localized state) broken translational symmetries.
\begin{figure}[htb]
    \centering
    \includegraphics[width=.8\linewidth]{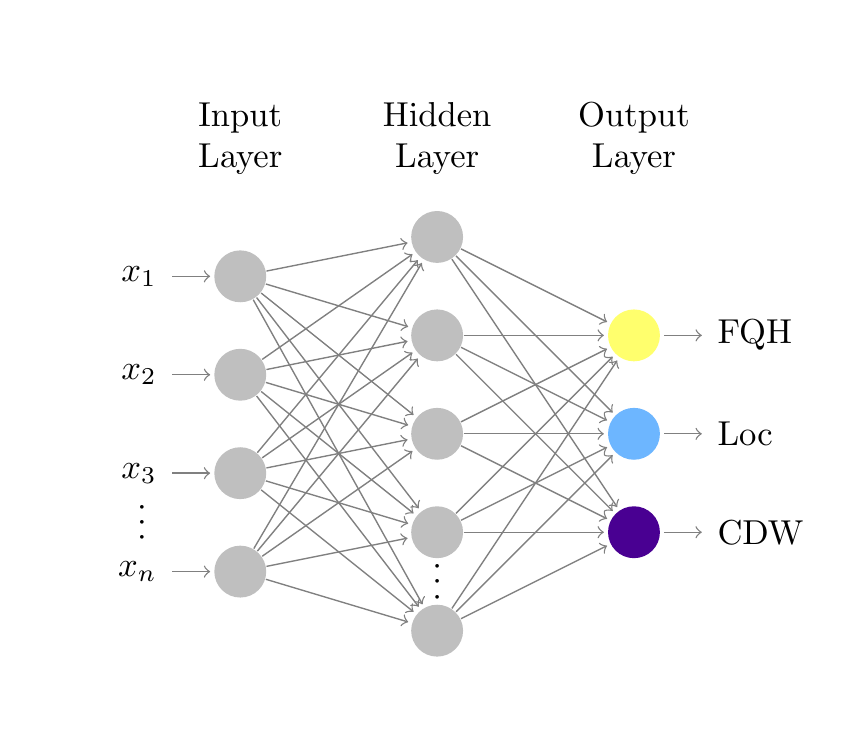}
    \caption{ Schematic diagram of neural network. We use the rank ordered entanglement spectrum or the charge density as input $x_i$. The hidden layer is specified by weights $w_{ij}$ and biases $b_i$.  We use a single hidden layer in the case of the rank ordered entanglement spectrum, and two in the case of the charge density. Each output neuron corresponds to a phase under consideration, with the output value the neural network's judgement on the likelihood of the phase. We normalize outputs using a softmax layer. }
    \label{fig:schematic_net}
\end{figure}

To obtain the ES and CD data, we exactly diagonalize the full Hamiltonian in Eq.~(\ref{eq:ham}) for each point in the two-dimensional parameter space $(W, \Delta V_1)$ and each disorder configuration.
From the pure many-body ground state of the system $|\psi\rangle$, we obtain the bi-partite ES, $\{ \xi_i \}$, by partitioning the system in the orbital space~\cite{li:prl2008a}
, as indicated in Fig.~\ref{fig:torus}. 
We note that, because of Gaussian localization of single-particle orbitals in a Landau level, the partitioning in orbital space roughly corresponds to an actual partition in real space.
 Having fixed the choice of the partition, we then 
perform the standard Schmidt decomposition $|\psi\rangle = \sum_i e^{-\xi_i/2} |A_i\rangle |B_i\rangle$, where the vectors $|A_i \rangle$ form an orthonormal basis for the subsystem $A$ (and similarly for subsystem $B$). 
Note that the von Neumann (entanglement) entropy can be directly computed from $\{\xi_i\}$ via the expression $S=-\sum_i \xi_i \ln \xi_i$. 
On the other hand, the charge density is defined as $\rho(\mathbf{x}) = \langle \psi | \hat{\Psi}^\dagger(\mathbf{x}) \hat{\Psi}(\mathbf{x}) |\psi\rangle$, where $\hat{\Psi}(\mathbf{x})$ is the electron field operator, and we evaluate the density 
on a $20\times 20$ mesh of evenly spaced points $\mathbf{x}$.

In one case, we use the obtained ES data as input to our ANN as a rank ordered list of 
numbers $\xi_i$. We do not consider reduced density matrix eigenvalues below numerical
precision, instead setting the corresponding $\xi_i$ to zero.  
In the other case, we use the CD data as input to our ANN as a vectorized two-dimensional
image.
The architecture of our choice is a 
fully-connected feed-forward artificial neural network
with a single hidden layer containing 50 neurons
for processing the ES data, and two hidden layers with 200 and 50
neurons for the CD data, see Fig.~\ref{fig:schematic_net}. 
Each neuron $j$ processes the inputs $x_i$ according to the weight matrix $w_{ji}$ and the bias vector $b_j$ specific to that neuron $ \sigma(w_{ji}x_i + b_j)$ where the rectified linear activation function is given by $\sigma(y)\equiv\rm{max}(y,0)$. The sum of the neural outputs
is normalized via a softmax layer.

Given the theoretical insight, we train the neural network using data from $5000$ disorder configurations at one or two training points inside each phase, as detailed below.
We use cross-entropy as the cost function for stochastic
gradient descent.
Once the network is trained to $>99\%$ accuracy, we fix the weights and biases and let the ANN recognize the phase associated with the rest of the phase space by averaging the neuron outputs from 500 disorder configurations for each phase space point.

\section{Results}\label{sec:results}
\begin{figure}[htb]
    \centering
    \includegraphics[width=\linewidth]{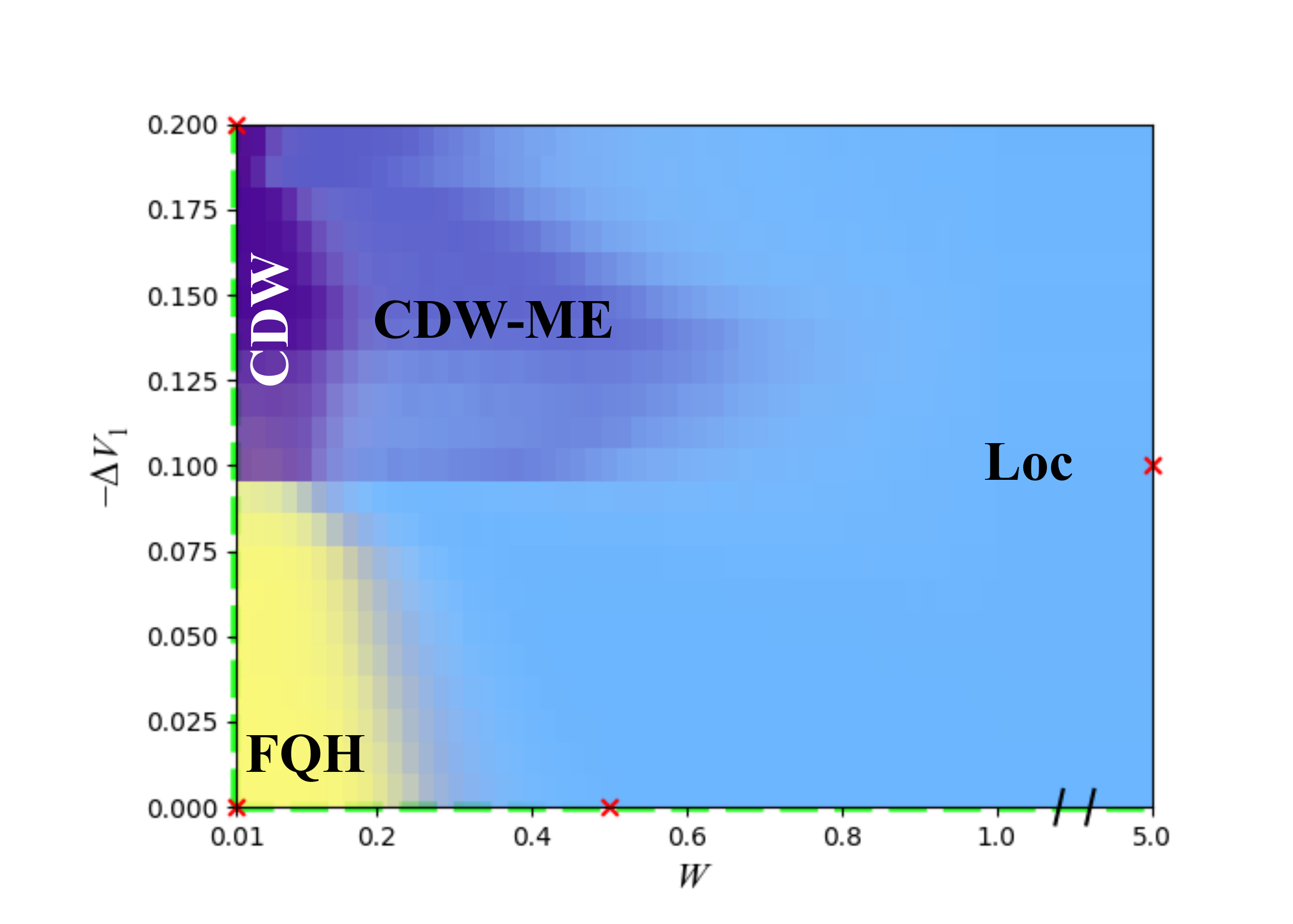}
    \caption{The phase diagram based on the output from three output-neuron neural network. Training points are marked with 
    a red cross on the diagram. Black slashes signify an axis break.
    Left and bottom side cuts (emphasized with green dashed lines) are shown in Figs.~\ref{fig:clean-cut} and
    \ref{fig:wscaling_plot}, respectively. Outputs of each neuron 
    scale from transparent to opaque as output goes from zero to one.
    We decrease the opacity of the output obtained using the ES by $50\%$, and layer it over the output obtained using the charge density.}
    \label{fig:multiphase_plot}
\end{figure}

Our multifaceted approach reveals new insight into the characteristics of topological, symmetry-breaking, and localized phases and their competition. The full phase diagram in shown in Fig.~\ref{fig:multiphase_plot}. 
This phase diagram was obtained for the system with $N=5$ electrons,
with NN training points indicated by red crosses. 
Although the lack of symmetry prevents one from reaching large system sizes, we note that 
qualitatively similar phase diagrams are obtained for other values of $N$. 
 The phase diagram contains four distinct regions that have been labelled in Fig.~\ref{fig:multiphase_plot}. In order to develop some intuition behind the identification of these phases, it is instructive to also inspect the typical (i.e. from a single, arbitrary disorder configuration) CD and ES in each of the regions in the phase diagrams, which are shown in  Figs.~\ref{fig:cd_profiles} and \ref{fig:es_profiles}, respectively. We next discuss in detail each of the four phases in the diagram, and their transitions to neighboring phases.

\subsection{Identification of phases}

\begin{figure}[htb]
    \centering
    \includegraphics[width=\linewidth]{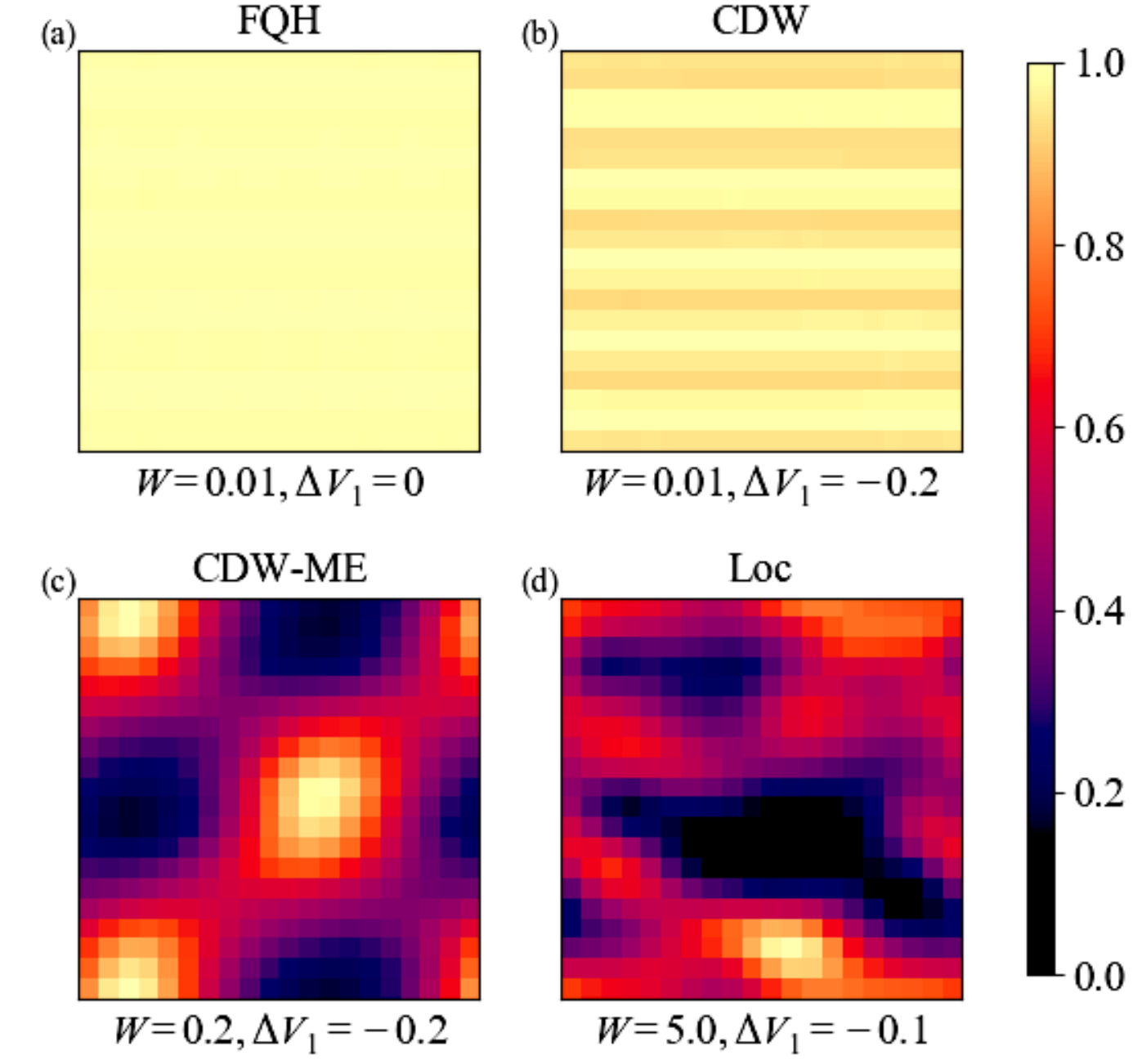}
    \caption{
        Typical real space charge density profiles $\rho(\mathbf{x})$ in (a) the FQH phase,
        (b) the CDW phase, (c) the intermediate CDW-ME state, and
        (d) the localized state. In each case the charge density is 
        plotted as a fraction of the maximum value it attains on the
        torus, and colorized as shown in the color bar on the right. 
        The charge density is largely uniform in the FQH phase, exhibits
        stripes in the CDW phase, separated "droplets" in the CDW-ME phase,
        and is seemingly random in the localized state. 
    }
    \label{fig:cd_profiles}
\end{figure}

First, we identify a robust region corresponding to topological order of the Laughlin $\nu=1/3$ state, indicated by the yellow color in Fig.~\ref{fig:multiphase_plot}. As the Laughlin state represents an incompressible liquid, we expect its CD to be spatially uniform, as indeed confirmed by Fig.~\ref{fig:cd_profiles}(a). Furthermore, as the Laughlin phase is a \emph{gapped} liquid, it remains stable for some finite amount of perturbations, either by softening of the interactions ($-|\Delta V_1|$) or by disorder $W$. For this region, the ANN output based on CD and ES perfectly match in identification of the phase boundary.
 
In the clean limit ($W\ll 1$), the system is expected to undergo a quantum phase transition from the FQH phase into a CDW phase upon increasing the magnitude of the $V_1$ pseudopotential perturbation $-|\Delta V_1|$~\cite{prange:s1990a}. 
Indeed, we can see in Fig.~\ref{fig:multiphase_plot}
that for weak disorder ($W=0.01$) and sufficiently negative $-|\Delta V_1| = 0.2$, 
the ANN finds a phase transition based on both CD and ES facets of the data. The plot of the charge density in the large $-|\Delta V_1|$ region Fig.~\ref{fig:cd_profiles}(a) clearly shows a stripe CDW phase.

We now turn to the light purple region of Fig.~\ref{fig:multiphase_plot} at intermediate disorder strength. At moderate disorder strength of $W_{c1}\lesssim 0.2$, the long-range CDW (i.e., correlation length of the order of the system size) is destroyed. Here the two facets of the data, CD and ES give us different insights. 
From the CD facet, the formation of large droptlets as shown in Fig.~\ref{fig:cd_profiles}(c) is recognized by the ANN to be similar to the CD distribution of localized phase in Fig.~\ref{fig:cd_profiles}(d). But clearly the distribution of charge in Fig.~\ref{fig:cd_profiles}(c) is more organized to the extent that it is 
somewhat reminiscent of a crystalline CDW bubble phase~\cite{bubble1,bubble2,bubble3}
that is usually discussed in the context of clean, fractionally filled higher Landau levels. 
However, considering the boundary conditions, the CD only exhibits two droplets and does not really match
the description of a crystalline state with multiple electrons per site.

\begin{figure}[htb]
    \centering
    \includegraphics[width=\linewidth]{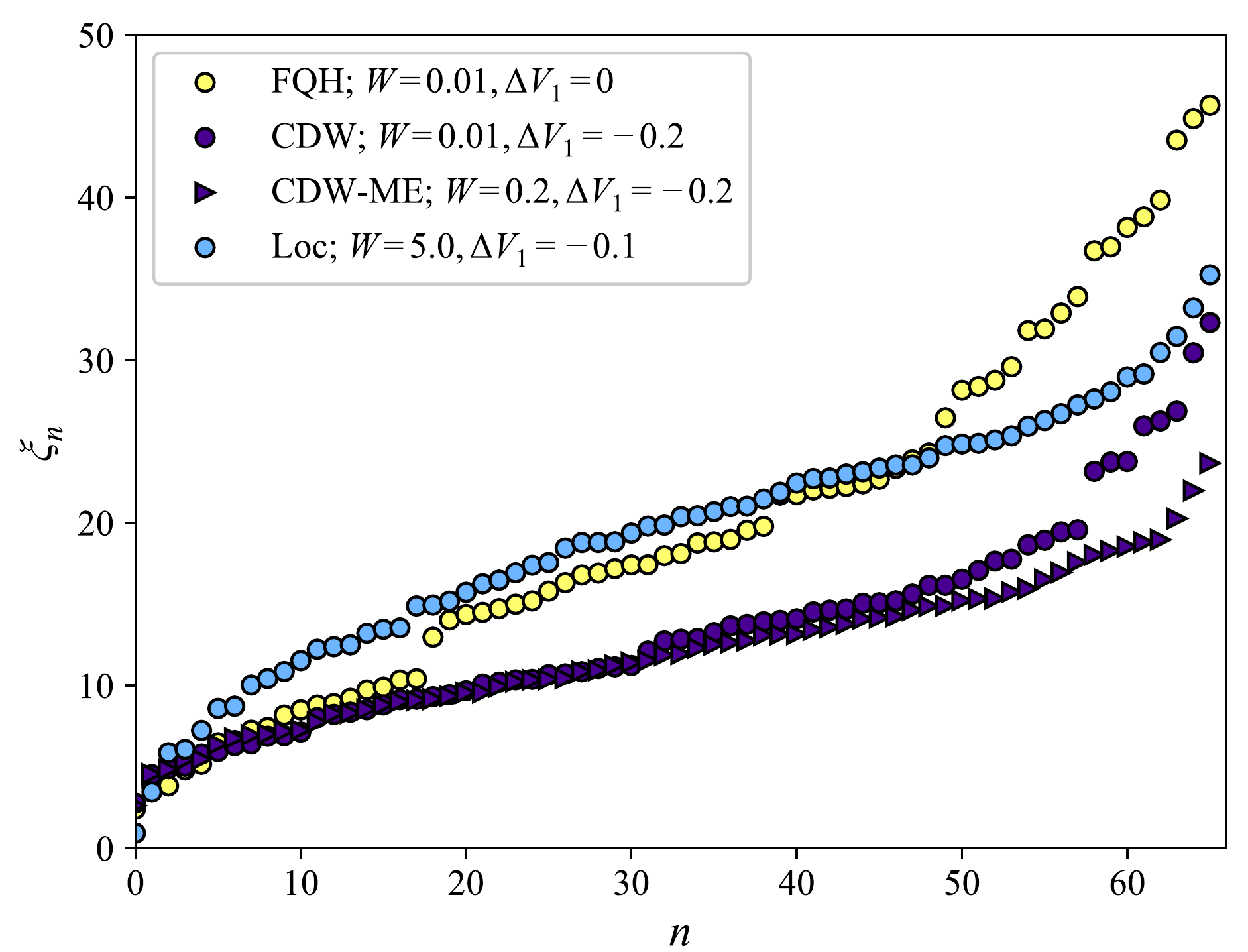}
    \caption{
     Typical entanglement spectra in each of the four phases identified in 
    Fig.~\ref{fig:multiphase_plot}. 
    The entanglement 
    energies $\xi_n$ are plotted against their indices $n$ in the
    rank-ordered entanglement spectrum. We observe that the ES are 
    visually distinct between the FQH, localized, and CDW states. 
    However, the CDW-ME looks comparatively similar to the CDW
    from the ES perspective, especially in the lower entanglement
    energy part of the ES.  }
    \label{fig:es_profiles}
\end{figure}

Further insight into this light purple region is afforded by looking at the ES facet. While the ANN assessment of the region based on the CD data was to identify it with the localized phase, this changes when the ES data is given. Actually, the ANN looking at the ES data identifies this region with the CDW phase. Looking at the plot of typical rank-ordered entanglement spectra shown in Fig.~\ref{fig:es_profiles} it is clear that the ES of this intermediate disorder regime is distributed in a fashion quite similar to that of the CDW phase, especially at low entanglement energies. Hence, the comparison between the ANN assessment based on CD and that based on ES is that this region shares characteristics of both a heterogeneous state and a CDW state.

According to the Imry-Ma argument\cite{PhysRevLett.35.1399}, the CDW order is expected to first break into droplets of CDW states of correlation length smaller than the system size. Although the stripe pattern is invisible in the CD distribution of this region in Fig.~\ref{fig:cd_profiles}(c) due to the resolution of our calculation, the fact that the size of the droplets are such that they can contain several CDW wavelengths makes it plausible that the purple region is supporting a  CDW-microemulsion (CDW-ME) rather than a featureless localized state. A new finding is that ANN can distinguish this state from  a localized state, which is realized at even stronger disorder 
[see Fig.~\ref{fig:cd_profiles}(d)], 
based on the ES facet.
Our ANN is finding the ES structure of this CDW-ME  to be identifiable as that of the long-range CDW state, while the CD structure of the CDW-ME to be identifiable as that of localized state. This identification of the CDW-ME is a new diagnostic afforded by our multifaceted application of ANN.

\subsection{Phase transitions}
We now examine in detail the transitions between the phases by studying one-parameter slices of our phase diagram in Fig.~\ref{fig:multiphase_plot}. This will provide us with further benchmarks against some results that are available in the literature, which have been obtained via more conventional diagnostics of quantum phase transitions varying only one of the two parameters of our phase space. 

\begin{figure}[htb]
    \centering
    \includegraphics[width=\linewidth]{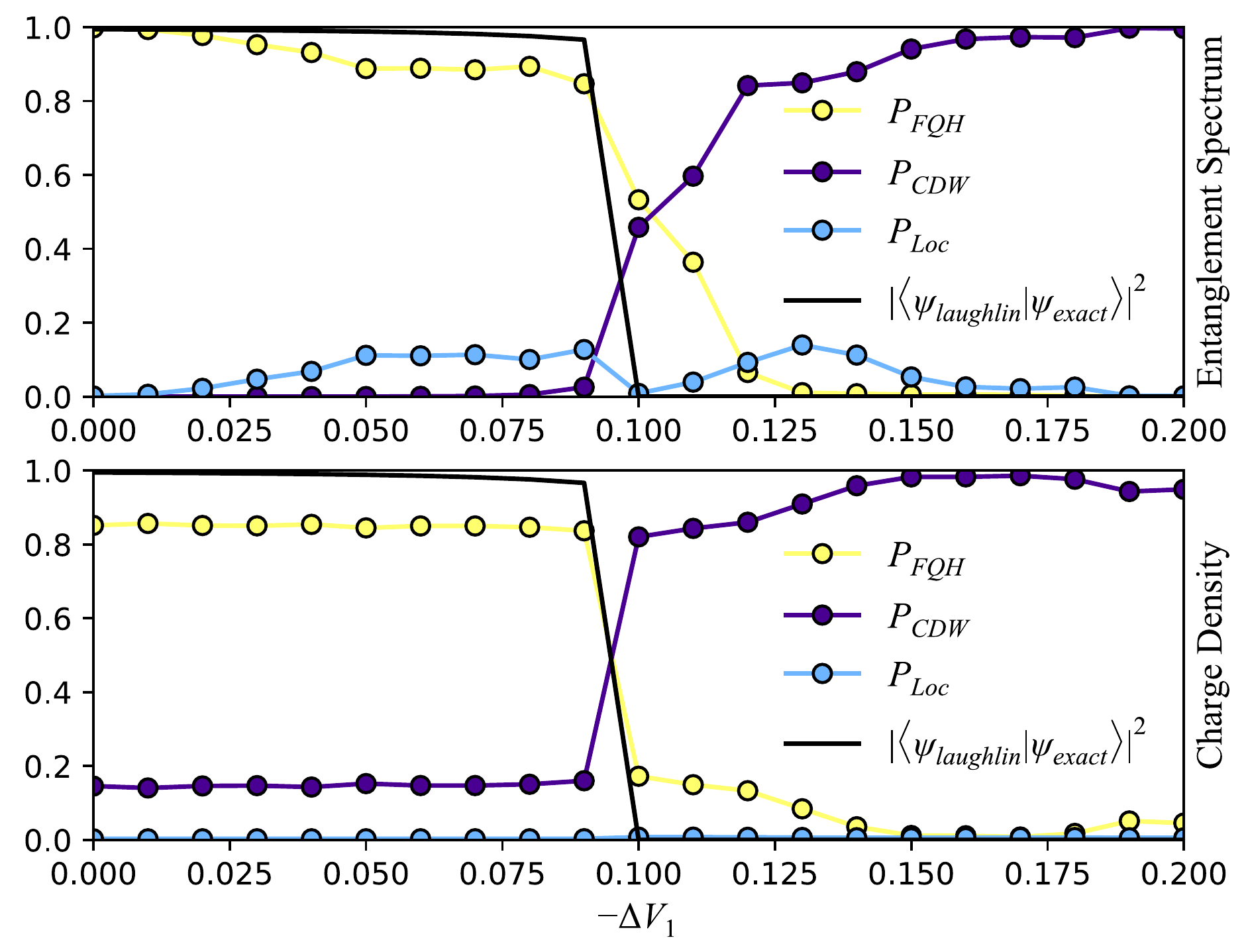}
    \caption{(a) The FQH-CDW transition along $W=0.01$. Neural network output for probabilities of the system being
    in the FQH, CDW, and localized states ($P_{FQH}$,$P_{CDW}$, and
    $P_{Loc}$, respectively) as a function of pseudopotential shift  $\Delta V_1$ for $N_e = 5$ using  the ES (upper panel) and the CD (lower panel) as input
    to the ANN. Phase transition to CDW is predicted around $\Delta V_1 = -0.1$. (b) The overlap between the Laughlin state and the exact ground state of the Hamiltonian. Precipitous drop in overlap also 
    predicts phase transition around $\Delta V_1 = -0.1$.
    }
    \label{fig:clean-cut}
\end{figure}

In the clean limit, a 
transition between a CDW phase and an FQH state is known to occur as a function of $\Delta V_1$~\cite{prange:s1990a}. 
The cut along the $\Delta V_1$ axis is shown in Fig.~\ref{fig:clean-cut}(a), and it indeed reveals a sharp transition between the FQH output dominant region and the CDW output dominant region around $-\Delta V_1 \approx 0.1$.
It is illuminating to contrast this neural network based detection of the phase transition to the conventional measures such as the wave function overlap with the Laughlin state, which is shown in Fig.~\ref{fig:clean-cut}(b). We observe that the overlap drops sharply to near zero at around the same value $-\Delta V_1 \approx 0.1$ as well. Note that because we still have very weak but non-zero disorder ($W=0.01$) in Fig.~\ref{fig:clean-cut}, we present the overlap between equal amplitude
superpositions of the three topologically degenerate Laughlin states and
the three lowest energy eigenstates of the exact Hamiltonian, which
are topologically degenerate in the FQH phase. 
The remarkable agreement between different diagnostics of the transition suggests that our neural network can accurately distinguish competing phases that are not related by symmetry, solely based on either the rank-ordered entanglement spectrum without reference to a good quantum number or the charge density. 

\begin{figure}[htb]
    \centering
    \includegraphics[width=\linewidth]{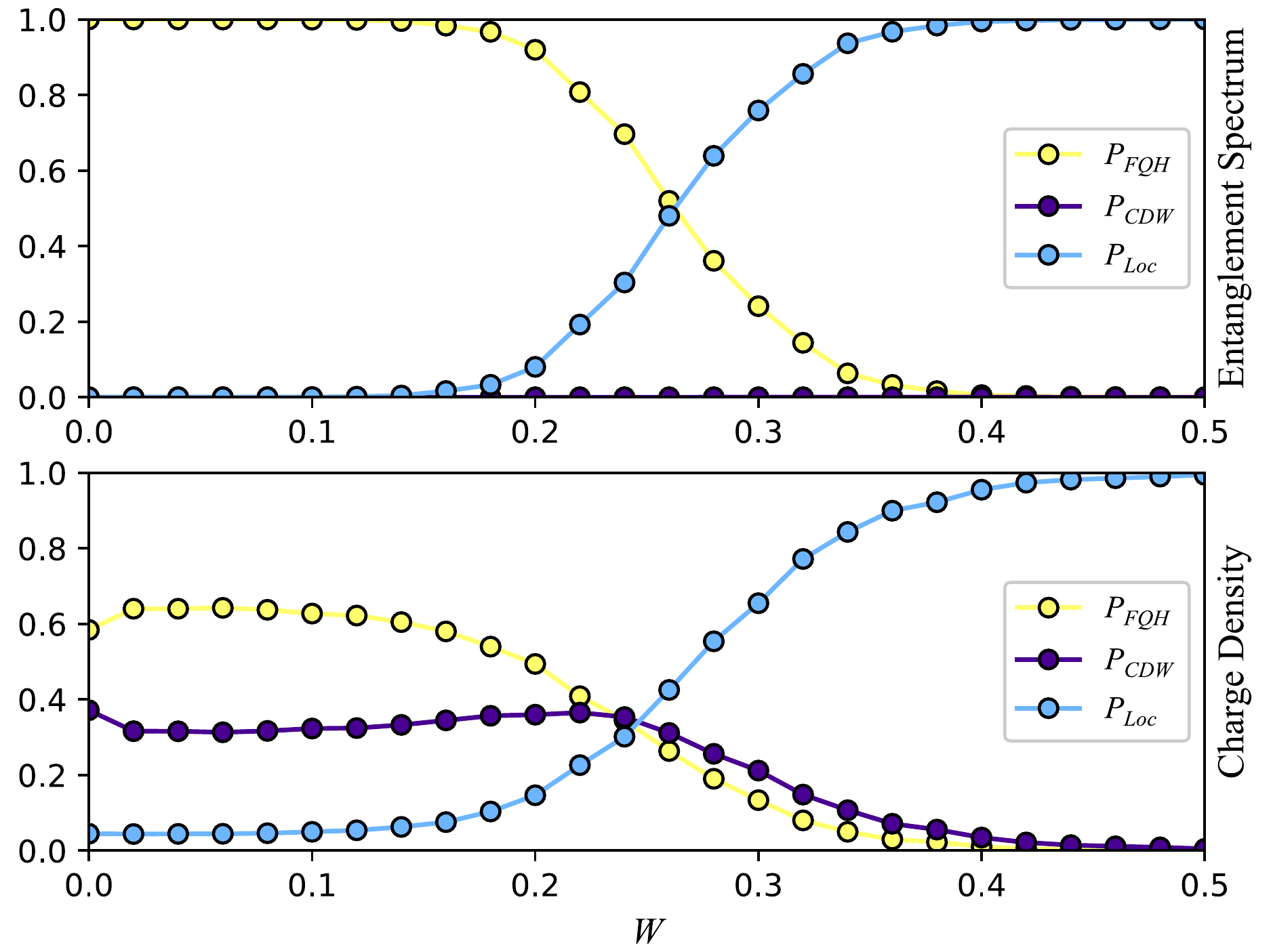}
    \caption{
    The FQH-localized state transition at $\Delta V_1 = 0$.
    Neural network output probabilities of the system
    being in the FQH, CDW, and localized states 
    ($P_{FQH}$, $P_{CDW}$, and $P_{Loc}$ respectively) as a function of 
    disorder strength $W$ for $N_e = 5$
    observed over $W \in [0.01,0.5]$ using (upper panel) the ES and (lower panel) CD as input to the ANN. Phase transition to the localized state is predicted around 
    $W = 0.26$. 
    }
    \label{fig:wscaling_plot}
\end{figure}

Finally, we turn to the localization transition along the $W$ axis and pure Coulomb interaction ($\Delta V_1=0$).
The $\Delta V_1=0$ cut shown in Fig.~\ref{fig:wscaling_plot} shows a broad transition around $W_c \approx 0.26$.  In the case of the
CD input (lower panel), we assess the threshold as in Ref.~\onlinecite{venderley:prl2018a}: at $W=0.26$, the output
from all three neurons is approximately $1/3$, corresponding to a point of maximal confusion
for a three neuron softmax layer.
Unlike the CDW transition in the clean case, however, there is no rigorously established conventional criteria for this localization transition. 
Liu and Bhatt~\cite{liu:pr2017a,liu:prl2016a} have recently proposed tracking the finite-size scaling of the numerical derivative of the ground state entanglement entropy with respect to disorder strength, $dS/dW$. This diagnostic (also used in  Ref.~\onlinecite{schliemann:pr2011a} for a bilayer quantum Hall system) puts the threshold at a much smaller disorder strength of $W\approx 0.09$. On the other hand, our threshold is consistent with the result found in Refs.~\onlinecite{sheng:prl2003a,wan:pr2005a} using the fluctuations in the total Chern number of the degenerate ground states on a torus which found $W_c \approx 0.22 \pm 0.025$. 

\section{Conclusions}\label{sec:conc}

We have used supervised machine learning via ANN to study the disorder-interaction phase diagram involving three competing states: FQH, CDW, and localized state.  
Using multiple facets of the data (rank-ordered ES and CD), we have reproduced known results along the two axes of the phase diagram. 
In the weak disorder limit ($W \lesssim 0.1$), the ANN finds the phase transition between the FQH phase and the CDW phase around $\Delta V_1 \approx -0.1$, which agrees with other estimates of the transition based on wave function overlap. In the Coulomb interaction limit ($\Delta V_1=0$), the ANN finds the FQH phase to localize 
 around disorder strength $W \approx 0.26$, roughly consistent with earlier findings using the total Chern number of the topologically degenerate ground states~\cite{sheng:prl2003a,wan:pr2005a}.

Furthermore, we have extended the previous results to the full two-dimensional $(W,-\Delta V_1)$ phase diagram. We found the FQH phase to be more robust to the disorder than CDW, as is expected from the fact that FQH is a topologically ordered state and CDW should be sensitive to disorder. At the same time, using the rank-ordered ES and CD as two independent facets of the computational data revealed the new regime of CDW-ME which has droplets of CDW with CDW correlation length less than the system size that nevertheless has the same entanglement structure as the CDW phase. 

Our study also shows that
the ES can serve as a diagnostic
of phase transitions in the case of competing interaction and disorder
when the relevant phases are distinguished by entanglement properties,
even in the absence of a traditional organizing principle for the ES.
Moreover we have seen that the ES contains structure that understands the formation
of multiple CDW droplets as being similar to a single CDW, as opposed to just recognizing symmetry breaking, which appears to be the only structure learned from the CD.

A distinct advantage of the ML method is its numerical efficiency: it only requires a large
number of disorder configurations at the points of neural network
training, whereas interpolation can be performed by averaging over far
fewer (e.g. 500 at each interpolation point
compared to several thousand at each training point).
There is also no need to search through the entirety of parameter space
to locate a phase transition: it is found directly by interpolating
between the neural network training points. This is especially 
advantageous in the case of a multidimensional parameter space. Note also that this method is not tied to any
particular system geometry or disorder model.
When studying the ground state of systems with competing interactions and 
disorder, the success of supervised
machine learning in studying transitions between the FQH, CDW, and
localized states supports the search for diagnostic quantities that 
distinguish the relevant phases that can be understood via machine
learning without needing to be understood by humans.

\section{Acknowledgements} 

We thank Jordan Venderley, Simon Trebst, Roger
G. Melko, Zhao Liu and Nicolas Regnault for useful discussions.  E-AK acknowledges DOE support under Award DE-SC0010313 and Award DOE	DE-SC0018946 as well as Simons Fellow in Theoretical
Physics Award $\#$392182. YZ acknowledge support from the Bethe
Postdoctoral Fellowship and from the W.M. Keck Foundation.
ZP acknowledges support by EPSRC grant EP/R020612/1. Statement of compliance with EPSRC policy framework on research data: This publication is theoretical work that does not require supporting research data.
The authors thank KITP supported by NSF grant NSF PHY-1748958, for its hospitality during the initial stage of the collaboration. 

\bibliographystyle{apsrev4-1}
\bibliography{library}

\end{document}